\documentclass[conference,a4paper]{APSIPA2025}
\usepackage{amsmath}
\usepackage{graphicx}
\usepackage{multirow}
\usepackage{amssymb,epsfig}
\usepackage{threeparttable}
\usepackage[backend=biber,style=ieee,]{biblatex}
\usepackage{geometry}
\usepackage{fancyhdr}
\fancypagestyle{firststyle}{
  \fancyhf{}
  \fancyhead[C]{2025 Asia Pacific Signal and Information Processing Association Annual Summit and Conference (APSIPA ASC)}
}

\begin{document}

\title{Speech Intelligibility Assessment with Uncertainty-Aware Whisper Embeddings and sLSTM}

\author{
\authorblockN{
Ryandhimas E. Zezario\authorrefmark{1} and
Dyah A.M.G. Wisnu\authorrefmark{2} and
Hsin-Min Wang\authorrefmark{3} and
Yu Tsao\authorrefmark{1} 
}

\authorblockA{
\authorrefmark{1}
Research Center for Information Technology Innovation, Academia Sinica, Taipei, Taiwan \\
E-mail: \{ryandhimas,yu.tsao\}@citi.sinica.edu.tw}

\authorblockA{
\authorrefmark{2}
Social Network and Human Centered Computing, Institute of Information Science, Academia Sinica, Taipei, Taiwan \\
E-mail: dyahayumgw@iis.sinica.edu.tw}

\authorblockA{
\authorrefmark{3}
Institute of Information Science, Academia Sinica, Taipei, Taiwan \\
E-mail: whm@iis.sinica.edu.tw}

}

\maketitle
\thispagestyle{firststyle}
\pagestyle{fancy}

\begin{abstract}
Non-intrusive speech intelligibility prediction remains challenging due to variability in speakers, noise conditions, and subjective perception. We propose an uncertainty-aware approach that leverages Whisper embeddings in combination with statistical features—specifically, the mean, standard deviation, and entropy computed across the embedding dimensions. The entropy, computed via a softmax over the feature dimension, serves as a proxy for uncertainty, complementing global information captured by the mean and standard deviation. To model the sequential structure of speech, we adopt a scalar long short-term memory (sLSTM) network, which efficiently captures long-range dependencies. Building on this foundation, we propose iMTI-Net, an improved multi-target intelligibility prediction network that integrates convolutional neural network (CNN) and sLSTM components within a multitask learning framework. It jointly predicts human intelligibility scores and machine-based word error rates (WER) from Google ASR and Whisper. Experimental results show that iMTI-Net outperforms the original MTI-Net across multiple evaluation metrics, demonstrating the effectiveness of incorporating uncertainty-aware features and the CNN-sLSTM architecture.
\end{abstract}

\section{Introduction}
Speech intelligibility is an essential indicator for evaluating a wide range of speech-related applications, including speech enhancement \cite{loizou2007speech, ref_19}, hearing aid (HA) devices \cite{katehaspi, kates2014hearingb}, and telecommunications \cite{ref_38}. The direct measurement is based on human listening tests, where listeners are required to recognize words from the played speech samples. The ratio of correctly recognized words to the total number of words is used to determine the intelligibility score. However, despite the reliability of human listening tests, a sufficient number of listeners is necessary to obtain unbiased measurements. Furthermore, this requirement limits the practicality and scalability of human evaluations. To overcome this problem, a series of signal processing-based approaches have been proposed, such as the articulation index (AI) \cite{ref_35}, speech intelligibility index (SII) \cite{ref_36}, extended SII (ESII) \cite{ref_37}, speech transmission index (STI) \cite{ref_38}, and short-time objective intelligibility (STOI) \cite{ref_39}. However, most of these methods require ground-truth references to produce reliable evaluation scores.

With the growing interest in reliable non-intrusive speech intelligibility methods, where ground-truth references are not necessarily required for evaluation, deep learning models have received increasing attention. This trend is further supported by the availability of large-scale datasets labeled by human annotators with corresponding speech intelligibility scores, enabling the development of deep learning-based intelligibility assessment models \cite{SIP_DL, andersen2018nonintrusive, ref_52, chiang2023multiobjective, tu22_interspeech, zezario2022mti, mosa-net, zezario2022mbi, cuervo2024speech, mogridge2024nonintrusive, zezario2024studyincorporatingwhisperrobust, MAWALIM2023109663}. In the early stages of this development, most well-known approaches \cite{SIP_DL, andersen2018nonintrusive, ref_52} relied on traditional speech processing techniques to extract acoustic features. More recently, the introduction of large-scale pre-trained speech models, such as self-supervised learning models (e.g., wav2vec \cite{wav2vec} and HuBERT \cite{hubert}) and weakly supervised models (e.g., Whisper \cite{Whisper}), has further advanced the field by providing richer and more generalizable acoustic features \cite{mosa-net, zezario2022mbi, cuervo2024speech, mogridge2024nonintrusive, zezario2024studyincorporatingwhisperrobust}. Despite these advancements, developing accurate and generalizable intelligibility assessment models remains challenging. The subjective nature of intelligibility, variations in speaker characteristics, and environmental conditions can degrade prediction performance. Additionally, capturing long-range dependencies in temporal speech patterns is non-trivial, particularly under noisy or mismatched conditions. 

To address these issues, we propose an uncertainty-aware approach that combines Whisper embeddings with statistical features. Specifically, we extract the mean, standard deviation, and entropy of the Whisper embeddings across the feature dimension. We hypothesize that the mean and standard deviation capture global characteristics of the embeddings over time, while the entropy—computed by applying a softmax over the embedding feature dimension—serves as a proxy for uncertainty in the embedding representation. To effectively model the hierarchical and sequential structure of speech, we adopt the scalar long short-term memory (sLSTM) network \cite{beck2024xlstm}. The sLSTM employs scalar gating mechanisms that optimizes the memory mixing while maintaining the ability to capture long-range dependencies in sequential data.

Building on this foundation, we further extend our work by proposing an improved multi-target intelligibility prediction model (iMTI-Net), which enhances the original MTI-Net framework by integrating uncertainty-aware features and a convolutional neural network (CNN) with sLSTM model. The proposed model employs a multi-task learning strategy to simultaneously predict both human and machine intelligibility scores. Machine intelligibility scores are represented by character error rate (CER) from two automatic speech recognition (ASR) systems, namely Google ASR \cite{ASR} and Whisper \cite{Whisper}, while human intelligibility scores include subjective listening test scores and objective metrics such as STOI. Experimental results demonstrate that iMTI-Net consistently outperforms the original MTI-Net across nearly all evaluation metrics, highlighting the effectiveness of incorporating uncertainty-aware features and the CNN-sLSTM architecture. Specifically, for intelligibility prediction, the iMTI-Net with CNN-sLSTM achieves the highest linear correlation coefficient (LCC) (0.7817) and Spearman’s rank correlation coefficient (SRCC) (0.7622). For Whisper CER prediction, the CNN-BLSTM variant obtains the highest LCC (0.8151), while CNN-sLSTM achieves the best SRCC (0.8222) and MSE (0.0360). In the case of Google CER prediction, CNN-sLSTM outperforms all other models with the highest LCC (0.8505), SRCC (0.8403), and lowest MSE (0.0312). Similarly, for STOI prediction, CNN-sLSTM achieves the best LCC (0.9051) and SRCC (0.9150), while CNN-BLSTM obtains the lowest mean square error (MSE) (0.0031). These results confirm that the use of sLSTM, in combination with CNN and statistical features, leads to consistent and significant improvements in both human and machine intelligibility prediction tasks.

\graphicspath{ {./images/} }
\begin{figure}[t]
\centering
\includegraphics[width=8.3cm]{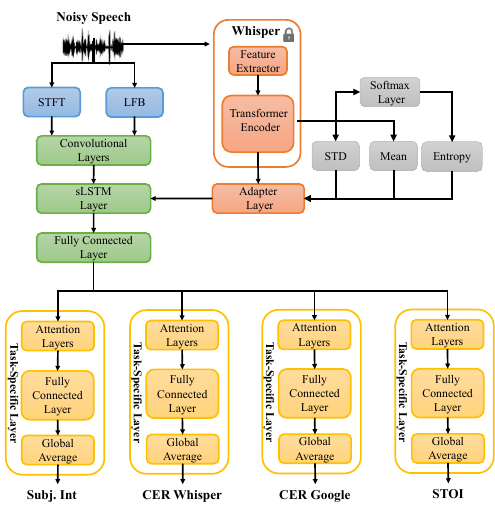} 
\caption{Architecture of iMTI-Net.} 
\label{fig:concat}
\end{figure}

\section{Proposed Method}

The overall architecture of the iMTI-Net model is illustrated in Fig.~1 and comprises three feature extraction modules. First, the speech waveform \(\boldsymbol{Y}\) undergoes short-time Fourier transform (STFT) to extract spectral features. Second, \(\boldsymbol{Y}\) is processed through learnable filter banks (LFB) within a sinc-based convolutional network \cite{sincnet}, producing complementary acoustic features. These two feature sets are concatenated along the feature dimension and fed into CNN layers; the output is denoted as \(\boldsymbol{C}\).

The third module processes the same waveform \(\boldsymbol{Y}\) using the Whisper model. Instead of using the Whisper embeddings directly as in prior work \cite{zezario2024studyincorporatingwhisperrobust}, our method extracts proxy uncertainty from these embeddings to provide richer information for intelligibility prediction. Specifically, we calculate the mean, standard deviation, and entropy of the embedding vectors at each time frame, which serve as proxies for the confidence and variability inherent in the embeddings. Formally, this is defined as:

\begin{equation}
\begin{array}{ll}
\boldsymbol{E} = \text{Whisper}(\boldsymbol{Y}) \in \mathbb{R}^{T \times D} \\[6pt]
\mu_t = \dfrac{1}{D} \sum\limits_{d=1}^{D} E_{t,d} \\[6pt]
\sigma_t = \sqrt{ \dfrac{1}{D} \sum\limits_{d=1}^{D} (E_{t,d} - \mu_t)^2 } \\[6pt]
\boldsymbol{p}_t = \mathrm{softmax}(\boldsymbol{E}_t) \\[6pt]
h_t = -\sum\limits_{d=1}^{D} p_{t,d} \log p_{t,d} \\[6pt]
\boldsymbol{x}_t = [\, \boldsymbol{E}_t ; \mu_t ; \sigma_t ; h_t \,] \in \mathbb{R}^{D+3} \\[6pt]
\tilde{\boldsymbol{x}}_t = [\, \operatorname{Adapter}(\boldsymbol{x}_t) ; \boldsymbol{C}_t \,]
\end{array}
\label{eq:whisper_features}
\end{equation}
where \(\boldsymbol{E} \in \mathbb{R}^{T \times D}\) denotes the Whisper embeddings with \(T\) frames and \(D\) dimensions. \(\mu_t\), \(\sigma_t\), and \(h_t\) represent the per-frame mean, standard deviation, and entropy, respectively, computed over the embedding dimensions. The feature vector \(\boldsymbol{x_t}\) is formed by concatenating the original embedding \(\boldsymbol{E}_t\) with its corresponding statistical features. Finally, \(\boldsymbol{\tilde{x}}_t\) denotes the final feature representation, formed by concatenating the output of the adapter layer applied to \(\boldsymbol{x}_t\) with the CNN-based acoustic feature \(\boldsymbol{C}_t\).

Next, the feature representation \(\boldsymbol{\tilde{x}}_t\) is processed through a sLSTM \cite{beck2024xlstm} network, which models temporal dependencies in the data. The sLSTM updates its internal state and output at each time step \(t\) based on the current input \(\boldsymbol{\tilde{x}}_t\) and the previous hidden state \(\boldsymbol{h}_{t-1}\). The computations are defined as follows:

\begin{equation}
\begin{array}{ll}
\boldsymbol{\tilde{z}}_t = \boldsymbol{w}_z^\top \boldsymbol{\tilde{x}}_t + \boldsymbol{r}_z \boldsymbol{h}_{t-1} + \boldsymbol{b}_z, & \boldsymbol{z}_t = \phi(\boldsymbol{\tilde{z}}_t) \\[6pt]

\boldsymbol{\tilde{i}}_t = \boldsymbol{w}_i^\top \boldsymbol{\tilde{x}}_t + \boldsymbol{r}_i \boldsymbol{h}_{t-1} + \boldsymbol{b}_i, & \boldsymbol{i}_t = \exp(\boldsymbol{\tilde{i}}_t) \\[6pt]

\boldsymbol{\tilde{f}}_t = \boldsymbol{w}_f^\top \boldsymbol{\tilde{x}}_t + \boldsymbol{r}_f \boldsymbol{h}_{t-1} + \boldsymbol{b}_f, & 
\boldsymbol{f}_t = 
\begin{cases}
\exp(\boldsymbol{\tilde{f}}_t) \\
\sigma(\boldsymbol{\tilde{f}}_t)
\end{cases} \\[8pt]

\boldsymbol{\tilde{o}}_t = \boldsymbol{w}_o^\top \boldsymbol{\tilde{x}}_t + \boldsymbol{r}_o \boldsymbol{h}_{t-1} + \boldsymbol{b}_o, & \boldsymbol{o}_t = \sigma(\boldsymbol{\tilde{o}}_t) \\[6pt]

\boldsymbol{c}_t = \boldsymbol{f}_t \cdot \boldsymbol{c}_{t-1} + \boldsymbol{i}_t \cdot \boldsymbol{z}_t \\[6pt]
\boldsymbol{n}_t = \boldsymbol{f}_t \cdot \boldsymbol{n}_{t-1} + \boldsymbol{i}_t \\[6pt]
\boldsymbol{\tilde{h}}_t = \dfrac{\boldsymbol{c}_t}{\boldsymbol{n}_t}, & \boldsymbol{h}_t = \boldsymbol{o}_t \cdot \boldsymbol{\tilde{h}}_t
\end{array}
\label{eq:slstm_clean}
\end{equation}
where, \(\boldsymbol{w}_*\), \(\boldsymbol{r}_*\), and \(\boldsymbol{b}_*\) are input weights, recurrent weights, and biases corresponding to each gate and cell units. The functions \(\phi(\cdot)\) and \(\sigma(\cdot)\) denote the hyperbolic tangent and sigmoid activation functions, respectively.

Furthermore, compared to the standard LSTM, the sLSTM maintains an additional normalization state \(\boldsymbol{n}_t\) alongside the cell state \(\boldsymbol{c}_t\), allowing it to compute a normalized hidden state \(\boldsymbol{\tilde{h}}_t = \frac{\boldsymbol{c}_t}{\boldsymbol{n}_t}\). This normalization helps stabilize training by preventing unbounded growth of the cell memory and improves the model’s ability to capture long-range dependencies efficiently. The output of the sLSTM is further processed by a fully connected layer. Task-specific layers are then employed to predict each corresponding intelligibility metric. Finally, the iMTI-Net framework integrates both frame-level and utterance-level scores into the objective function for each metric's loss.

\begin{equation}
\label{eq:loss}
   \small
    \begin{array}{c}
   L = \gamma_{1}L_{Int.} + \gamma_{2}L_{CER_{ws}}
 + \gamma_{3}L_{CER_{goo}}
  + \gamma_{4}L_{STOI}
    \end{array} 
\end{equation}
where weights between Intelligibility, CER Whisper, CER Google, and STOI are determined by $\gamma_{1}$, $\gamma_{2}$, $\gamma_{3}$, and $\gamma_{4}$, respectively. 

\section{Experiments}
\subsection{Experimental setup}
The iMTI-Net model was evaluated using the TMHINT-QI(S) dataset \cite{zezario2024studyincorporatingwhisperrobust}, an extended version of the original TMHINT-QI corpus \cite{TMINT-QI}. This extension incorporates additional unseen noise types, speakers, and speech enhancement models. Notably, TMHINT-QI(S) also serves as one of the benchmark tracks in the VoiceMOS Challenge 2023 \cite{cooper2023voicemos}. The evaluation set consists of clean, noisy, and enhanced speech samples, including three seen noise types (babble, white, and pink) and one unseen noise condition (street noise). It also covers three seen enhancement systems—minimum mean square error (MMSE) \cite{mmse}, fully convolutional network (FCN) \cite{FCN}, and transformer \cite{Trans}—alongside two unseen systems: conformer-based metric generative adversarial network (CMGAN) \cite{cao22_interspeech} and DEMUCS \cite{défossez2021music}. In total, the evaluation set consists of 1,960 utterances with a quality score (ranging from 0 to 5) and an intelligibility score (ranging from 0 to 1). Additionally, we prepared CER scores from two ASR systems—Google ASR \cite{ASR} and Whisper \cite{Whisper}. For consistency across metrics, the CER scores are inverted so that higher scores reflect better recognition performance.

To evaluate prediction performance, we adopt three commonly used metrics: LCC, SRCC \cite{srcc}, and MSE. Both LCC and SRCC assess the strength of the relationship between the predicted and ground-truth scores, with higher values indicating better alignment and overall performance. On the other hand, lower MSE values indicate better performance.

\subsection{Performance comparison between baseline and iMTI-Net}
In this experiment, we compare the performance of iMTI-Net with the original MTI-Net \cite{zezario2022mti}. While we follow the original implementation of MTI-Net for model deployment, we introduce slight modifications by incorporating additional assessment metrics during training. The original MTI-Net was trained to predict Intelligibility, Google CER, and STOI. In our setup, we also include Whisper CER as an additional target. Furthermore, given the strong performance of the Whisper model, we replace HuBERT with Whisper in our implementation. For simplicity, we refer to this modified version of MTI-Net as the baseline.

For the iMTI-Net model, we deploy two variants: one using a CNN-BLSTM architecture and the other using a CNN-sLSTM architecture. The CNN-BLSTM model consists of 12 convolutional layers with four channel groups (16, 32, 64, and 128 channels), followed by a one-layer BLSTM with 128 units and a fully connected layer with 128 neurons. Four prediction branches are used, each comprising an attention layer, a fully connected layer with a single output neuron, and a global average pooling operation to generate the predicted quality and intelligibility scores. During training, we set the loss weights as $\gamma_1 = 1$, $\gamma_2 = 1$, $\gamma_3 = 1$, and $\gamma_4 = 5$, and use a learning rate of 1e-5. Unlike the baseline model, which concatenates features along the temporal dimension, our iMTI-Net performs feature concatenation along the feature dimension. Additionally, the corresponding statistical features are concatenated with the original features in all iMTI-Net model variants. Lastly, the CNN-sLSTM version follows the same configuration, except that the sLSTM replaces the BLSTM component, while retaining the same architectural structure.

\begin{table}[t]
\caption{LCC, SRCC, and MSE results between Baseline and iMTI-Net for speech intelligibility prediction.}
\footnotesize
\begin{center}
 \begin{tabular}{c c c c c} 
 \hline
 \hline
 &  \textbf{Architecture} &\textbf{LCC} & \textbf{SRCC} & \textbf{MSE}  \\ [0.5ex] \cline{2-5}
 \hline\hline
Baseline&CNN-BLSTM&0.7630&0.7071&0.0249\\
iMTI-Net&CNN-BLSTM&0.7791&0.7581&0.0262\\
iMTI-Net&CNN-sLSTM&\textbf{0.7817}&\textbf{0.7622}&\textbf{0.0259}\\\hline
 \hline
\end{tabular}
\end{center}
\end{table}

\begin{table}[t]
\caption{LCC, SRCC, and MSE results between Baseline and iMTI-Net for CER of Whisper prediction.}
\footnotesize
\begin{center}
 \begin{tabular}{c c c c c} 
 \hline
 \hline
 &  \textbf{Architecture} &\textbf{LCC} & \textbf{SRCC} & \textbf{MSE}  \\ [0.5ex] \cline{2-5}
 \hline\hline
Baseline&CNN-BLSTM&0.7118&0.6600&0.0471\\
iMTI-Net&CNN-BLSTM&\textbf{0.8151}&0.8145&0.0334\\
iMTI-Net&CNN-sLSTM&0.8031&\textbf{0.8222}&\textbf{0.0360}\\\hline
 \hline
\end{tabular}
\end{center}
\end{table}

\begin{table}[t]
\caption{LCC, SRCC, and MSE results between Baseline and iMTI-Net for CER of Google prediction.}
\footnotesize
\begin{center}
 \begin{tabular}{c c c c c} 
 \hline
 \hline
 &  \textbf{Architecture} &\textbf{LCC} & \textbf{SRCC} & \textbf{MSE}  \\ [0.5ex] \cline{2-5}
 \hline\hline
Baseline&CNN-BLSTM&0.8156&0.7391&0.0373\\
iMTI-Net&CNN-BLSTM&0.8443&0.8358&0.0323\\
iMTI-Net&CNN-sLSTM&\textbf{0.8505}&\textbf{0.8403}&\textbf{0.0312}\\\hline
 \hline
\end{tabular}
\end{center}
\end{table}

\begin{table}[ht]
\caption{LCC, SRCC, and MSE results between Baseline and iMTI-Net for STOI prediction.}
\footnotesize
\begin{center}
 \begin{tabular}{c c c c c} 
 \hline
 \hline
 &  \textbf{Architecture} &\textbf{LCC} & \textbf{SRCC} & \textbf{MSE}  \\ [0.5ex] \cline{2-5}
 \hline\hline
Baseline&CNN-BLSTM&0.8693&0.8893&0.0060\\
iMTI-Net&CNN-BLSTM&0.9005&0.9050&\textbf{0.0031}\\
iMTI-Net&CNN-sLSTM&\textbf{0.9051}&\textbf{0.9150}&0.0039\\\hline
 \hline
\end{tabular}
\end{center}
\end{table}

Experimental results confirm that the proposed iMTI-Net consistently outperforms the Baseline across all evaluation metrics. As shown in Table 1, both iMTI-Net variants yield better performance for intelligibility prediction, with the CNN-sLSTM model achieving the highest LCC (0.7817) and SRCC (0.7622), along with a competitive MSE (0.0259). This highlights the effectiveness of combining statistical features with the sLSTM architecture in modeling intelligibility-related patterns. For Whisper CER prediction (Table 2), the iMTI-Net with CNN-BLSTM achieves the highest LCC (0.8151), while the CNN-sLSTM variant delivers the best SRCC (0.8222) and MSE (0.0360). In Table 3, the iMTI-Net with CNN-sLSTM again leads across all metrics for Google CER prediction—LCC (0.8505), SRCC (0.8403), and MSE (0.0312)—indicating that sLSTM offers a more effective sequential modeling framework than BLSTM in this context. Table 4 shows the results for STOI prediction, where iMTI-Net with CNN-sLSTM achieves the highest LCC (0.9051) and SRCC (0.9150), while iMTI-Net with CNN-BLSTM obtains the lowest MSE (0.0031). These findings further demonstrate the robustness and effectiveness of the proposed iMTI-Net, particularly with the CNN-sLSTM configuration.

\subsection{Qualitative analysis between baseline and iMTI-Net}

In this experiment, we focus on a qualitative comparison between the Baseline and iMTI-Net models. For iMTI-Net, we select the best-performing variant that uses CNN-sLSTM. The scatter plots, as shown in Fig.2, show that the Baseline tends to cluster predictions in the mid-range and struggles to predict very low or high intelligibility scores. In contrast, iMTI-Net produces a more even spread of predictions across the full range, showing better sensitivity to both low and high intelligibility. This suggests that the CNN-sLSTM structure and uncertainty-awareness help the model better capture variations in speech, leading to more accurate predictions.

\graphicspath{ {./images/} }
\begin{figure}[t]
\centering
\includegraphics[width=5.5cm]{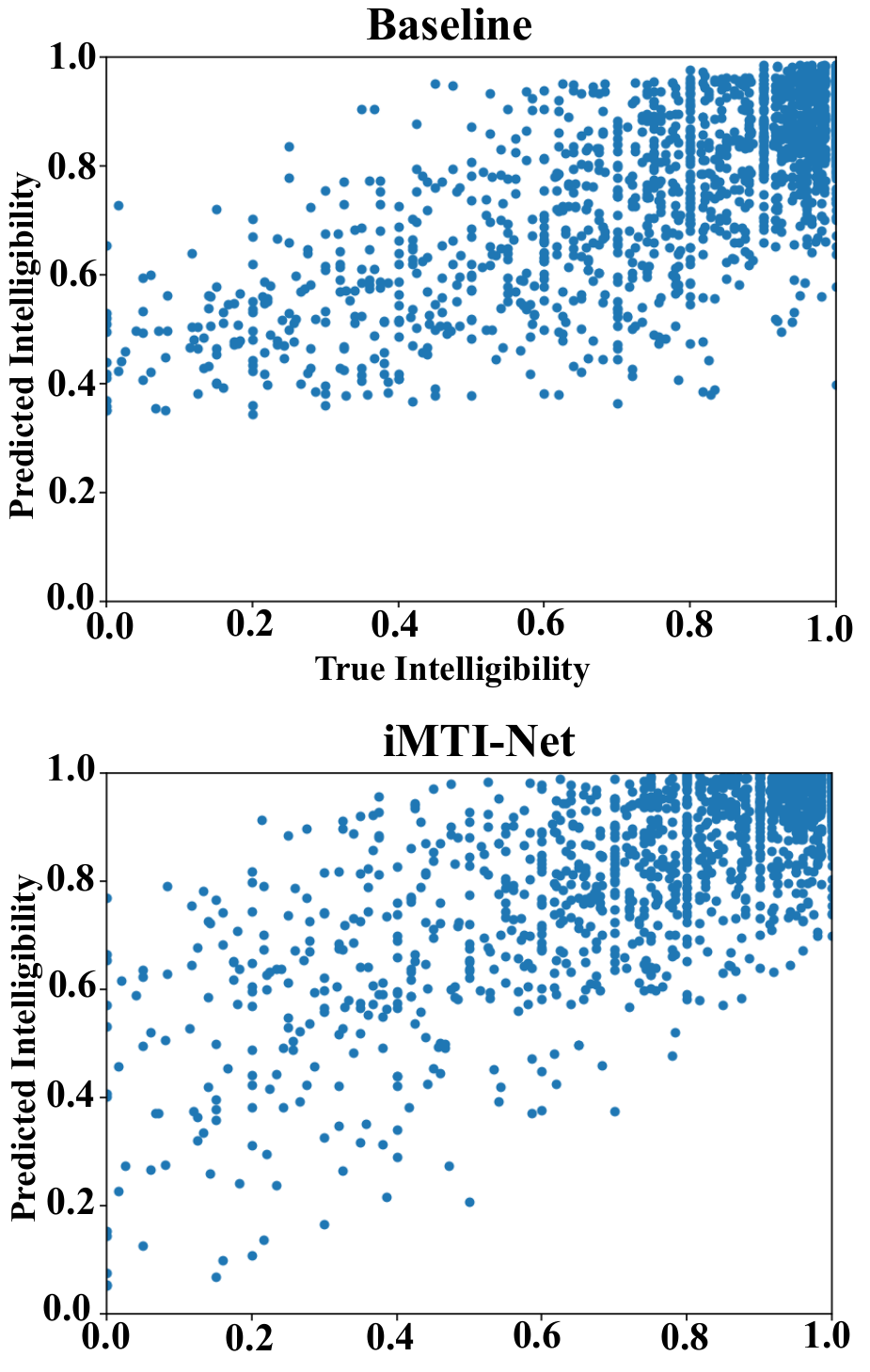} 
\caption{Scatter plots of Baseline and iMTI-Net for predicting subjective intelligibility.} 
\label{fig:concat}
\end{figure}

\section{Conclusions}
This work presents iMTI-Net, an improved multi-target model for non-intrusive speech intelligibility prediction. iMTI-Net integrates uncertainty-aware statistical features derived from Whisper embeddings with a CNN-sLSTM architecture and is trained using a multi-task learning strategy to jointly predict human intelligibility scores and machine-based CER from Google ASR and Whisper. To capture global speech characteristics and proxy uncertainty, we incorporate the mean, standard deviation, and entropy of the Whisper embeddings. Experimental results confirm that iMTI-Net consistently outperforms the original MTI-Net across intelligibility, STOI, and CER prediction tasks. The iMTI-Net with CNN-sLSTM variant achieves the best performance in most metrics, confirming the benefit of sLSTM in optimizing memory mixing while maintaining the ability to capture long-range dependencies in sequential data. In future work, we plan to explore broader use of uncertainty-aware representations and extend iMTI-Net to handle more diverse and unseen speech scenarios.

\printbibliography[title=References]
\end{document}